\documentclass[sensors,article,accept,pdftex,moreauthors]{Definitions/mdpi} 

\usepackage{comment}

\firstpage{1} 
\pubvolume{24}
\issuenum{5}
\articlenumber{1695}
\pubyear{2024}
\copyrightyear{2024}
\externaleditor{Academic Editor: Carles Gomez}
\datereceived{6 February 2024} 
\daterevised{4 March 2024} % Comment out if no revised date
\dateaccepted{4 March 2024} 
\datepublished{6 March 2024} 
\hreflink{https://doi.org/10.3390/s24051695} % If needed use \linebreak
\usepackage[style=long,nopostdot,acronym]{glossaries}
\makeglossaries
\renewcommand{\glossarysection}[2][]{}
\newacronym{API}{API}{Application Programming Interface}
\newacronym{CKAN}{CKAN}{Comprehensive Knowledge Archive Network}
\newacronym{CB}{CB}{Context Broker}
\newacronym{DCAT}{DCAT}{Data Catalog Vocabulary}
\newacronym{DCAT-AP}{DCAT-AP}{Data Catalog Vocabulary-Application Profile}
\newacronym{DET}{DET}{Data Enrichment Toolchain}
\newacronym{DQV}{DQV}{Data Quality Vocabulary}
\newacronym{EDP}{EDP}{European Data Portal}
\newacronym{EIF}{EIF}{European Interoperability Framework}
\newacronym{EOSC}{EOSC}{European Open Science Cloud}
\newacronym{ETSI}{ETSI}{European Telecommunications Standards Institute}
\newacronym{FAIR}{FAIR}{Findability, Accessibility, Interoperability and Reusability}
\newacronym{FCB}{FCB}{Federated Context Broker}
\newacronym{HTTP}{HTTP}{HyperText Transfer Protocol}
\newacronym{ICT}{ICT}{Information and Communication Technology}
\newacronym{IoT}{IoT}{Internet of Things}
\newacronym{JSON}{JSON}{JavaScript Object Notation}
\newacronym{JSON-LD}{JSON-LD}{JSON for Linked Data}
\newacronym{URL}{URL}{Uniform Resource Locator}
\newacronym{MQA}{MQA}{Metadata Quality Assessment}
\newacronym{NGSI-LD}{NGSI-LD}{Next Generation Service Interface with Linked Data}
\newacronym{OAI-PMH}{OAI-PMH}{Open Archives Initiative Protocol for Metadata Harvesting}
\newacronym{ODDM}{ODDM}{Open Data Development Model}
\newacronym{RDF}{RDF}{Resource Description Framework}
\newacronym{RDF/XML}{RDF/XML}{XML syntax for RDF}
\newacronym{SHACL}{SHACL}{Shapes Constraint Language}
\newacronym{SALTED}{SALTED}{Situation-Aware Linked heTerogeneous Enriched Data}
\newacronym{W3C}{W3C}{World Wide Web Consortium}
\newacronym{XML}{XML}{eXtensible Markup Language}

\Title{A Connector for Integrating NGSI-LD Data into Open Data~Portals}

\TitleCitation{A Connector for Integrating NGSI-LD Data into Open Data~Portals}

\Author{{Laura Mart\'in *\orcidA{},} Jorge Lanza \orcidB{}, V\'ictor Gonz\'alez \orcidC{}, Juan Ram\'on Santana \orcidD{}, Pablo Sotres \orcidE{} and Luis S\'anchez *\orcidF{}}

\AuthorNames{Laura Martín, Jorge Lanza, Víctor González, Juan Ramón Santana, Pablo Sotres and Luis Sánchez}

\AuthorCitation{{Martín, L.;} Lanza, J.; González, V.; Santana, J.R.; Sotres, P.; Sánchez, L.}
\address[1]{Network Planning and Mobile Communications Laboratory, Universidad de Cantabria, Plaza de la Ciencia s/n, 39005 Santander, Spain; jlanza@tlmat.unican.es (J.L.); vgonzalez@tlmat.unican.es (V.G.); jrsantana@tlmat.unican.es (J.R.S.); psotres@tlmat.unican.es (P.S.)
 \\
}

\corres{\hangafter=1 \hangindent=1.05em \hspace{-0.82em} Correspondence: lmartin@tlmat.unican.es (L.M.);  lsanchez@tlmat.unican.es (L.S.)}

\abstract{{Nowadays, there are plenty of data sources generating massive amounts of information that, combined with novel data analytics frameworks, are meant to support optimisation in many application domains. Nonetheless, there are still shortcomings in terms of data discoverability, accessibility and interoperability. Open Data portals have emerged as a shift towards openness and discoverability. However, they do not impose any condition to the data itself, just stipulate how datasets have to be described. Alternatively, the NGSI-LD standard pursues harmonisation in terms of data modelling and accessibility. This paper presents a solution that bridges these two domains (i.e., Open Data portals and NGSI-LD-based data) in order to keep benefiting from the structured description of datasets offered by Open Data portals, while ensuring the interoperability provided by the NGSI-LD standard. Our solution aggregates the data into coherent datasets and generate high-quality descriptions, ensuring comprehensiveness, interoperability and accessibility. The proposed solution has been validated through a real-world implementation that exposes IoT data in NGSI-LD format through the European Data Portal (EDP). Moreover, the results from the Metadata Quality Assessment that the EDP implements, show that the datasets’ descriptions generated achieve excellent ranking in terms of the Findability, Accessibility, Interoperability and Reusability (FAIR) data principles.}}

\keyword{NGSI-LD; Open Data; CKAN; DCAT-AP; interworking} 

\begin{document}

\section{Introduction}\label{sec:Introduction}
Proliferation of data sources is creating an abundance of information that is called to bring benefits for both the private and public sectors, increasing, for~example, administrations’ transparency and availability or fostering efficiency of public services~\cite{lourencco2017}. The~influence of \gls{ICT} on public values, particularly in bolstering transparency, has been acknowledged since the inception of computers in public agencies and institutions. This trend gained momentum with the widespread availability of Internet access in developed countries, intensifying competition in information supply and significantly easing the dissemination of~information.

In the contemporary landscape, global technological advancements heavily rely on data, and~the utilisation of \gls{ICT} in the public sector offers multifaceted benefits. Since 2010, the~concept of Open Data has been intricately linked to broader open government reforms, creating a connection between Open Data and reforms in public management~\cite{noveck2017}. For~example, Open Government Data has become a pivotal concept guiding government initiatives to enhance transparency and accountability regarding the utilisation of public resources. However, the~Open Data movement goes much further than openly disclosing data from public sectors. It has emerged as a catalyst for making data accessible to the public on the Internet. Increasingly, data from governmental, public, but~also private entities is being released online, primarily through Open Data portals. This extends to various sectors and contributes to the progression toward smart cities~\cite{pereira2017}.

Nonetheless, as~the volume of published resources grows, concerns arise regarding the quality and interoperability of the data sources and their corresponding metadata. These issues pose challenges to the usability of the resources. The~main software frameworks used for this kind of portals, like \gls{CKAN}, Socrata or OpenDataSoft, focus on the description of the datasets available through the portal, this is, on~the so-called metadata, while the actual data does not have to comply with any specific standard. It is usual that, even within the same portal, different datasets containing the same kind of data are hardly interoperable as they do not follow any specific information or data model. This situation gets worse the more inter-domain data the Open Data Portal stores. For~example, the~\gls{EDP} that harvests data from other portals all over Europe, or~National Data Portals that gather data from different Regional or Municipal~Governments.

This is the main limitation that the work that is described in this paper is tackling. It is critical that the data that can be retrieved through Open Data portals adheres to standard information and data models so that, besides~openly discoverable and accessible, it is actually usable and re-usable, without~facing tough interoperability challenges. Open Data portals, and~the technologies on which they are based, focus on the metadata, specifically on the datasets’ descriptions leaving the features of the actual data totally up to the data~provider.

In this sense, heterogeneity in data access stands as a significant obstacle impeding the broad usability of data, hindering the implementation of solutions that depend on data from various sources. Hence, ensuring data interoperability, encompassing both access and modelling, becomes imperative to facilitate the development of adaptable and reproducible solutions. To~achieve data interoperability, consensus is required on the technological interfaces and data modelling employed in data~exchange.

In this context, the~\gls{NGSI-LD} standard emerges as a potential solution to harmonise data access specifications, promoting interoperability among different data providers and consumers. \gls{NGSI-LD}, an~\gls{ETSI} standard, offers a comprehensive specification for enabling context data management. It can enhance access to context information by defining an \gls{API}~\cite{NGSI-LD-API} and an information model~\cite{NGSI-LD-InformationModel} for diverse participants. This standard is the core interface of the FIWARE open-source ecosystem. \gls{NGSI-LD} is already in use in numerous real-world pilots~\cite{teixeira2023, sotres2019}, providing a flexible and reliable means to address the challenges of data interoperability in scenarios where harmonising access to data from heterogeneous sources is~essential.

However, the~opposite situation to the one described above as the main limitation addressed by our work is faced in case the solution is solely based on \gls{NGSI-LD}. Its \gls{API} can be employed by applications to consume data, but~it does not offer native tools or frameworks to promote discoverability of the datasets that are accessible through such \gls{API}. Thus, \gls{NGSI-LD}, on~its own, cannot be the sole solution in terms of providing the Findability, Accessibility, Interoperability and Reusability (FAIR) data~principles.

The main novelty that we are bringing with the work that is described in the paper is to bring together the abovementioned two technologies, namely, \gls{CKAN}-based Open Data portals employing \gls{DCAT-AP} specification for describing their datasets, and~\gls{NGSI-LD}-based data, so that they can mutually cover each other limitations and realise a solution maximizing the Findability, Accessibility, Interoperability and Reusability (\gls{FAIR}) data principles. The~proposed solution eases the process for data providers managing their data following the \gls{NGSI-LD} standard to generate the corresponding metadata (i.e.,~datasets’ description) that is necessary to automatically expose their data on a \mbox{\gls{CKAN}-based} Open Data portals. Considering the great significance that \gls{FAIR} principles have been acquiring in the framework of data sharing, developing a solution that contributes to the maximization of such principles paves the way for establishing larger critical mass around well-established data and metadata modelling technologies (i.e., \gls{NGSI-LD} and \gls{DCAT-AP}, respectively) that should help in harmonizing the nowadays barely interoperable data exchange ecosystem. To~the best of our knowledge there are no similar alternatives proposed in the~literature.

In this paper, we are promoting the synergies that can be established among the Open Data portals, in~particular those based on \gls{CKAN} using the \gls{DCAT-AP} specification for describing the datasets that they contain, for~data availability and discoverability, and~the \gls{NGSI-LD} standard for data interoperability. In~this sense, this article is presenting the design and development of the solution, comprising a set of connectors that bridge these two domains. This solution is able to render \gls{NGSI-LD} data accessible through Open Data portals by aggregating the data into coherent datasets and generate high-quality descriptions, ensuring comprehensiveness, interoperability and~accessibility. 

The key idea is to integrate the capacity that Open Data portals have to describe the datasets as a whole, by~the use of the well-established \gls{W3C} \gls{DCAT-AP} specification~\cite{dcat-ap_specification}, and~the ability of providing the actual data (i.e., the data items within the available datasets) using a standard, interoperable and extendable information model through the adoption of \gls{NGSI-LD} syntax and semantics. This integration not only enriches the publicly available information but also expands the potential for insightful analysis and informed decision-making, aligning the worlds of IoT-generated data and open public~repositories.

The remaining of the paper is structured as follows. In~Section~\ref{sec:SotA} a brief review of key aspects of \gls{NGSI-LD} and Open Data is made, together with a discussion on related works currently available in the literature. Section~\ref{sec:Connector:Phase0} showcases the binding among the several data formats used throughout the architecture. The~\gls{NGSI-LD}-to-\gls{CKAN} connector design and implementation details are described in Section~\ref{sec:Connector}. In~Section~\ref{sec:Validation} the practical validation that has been carried out to assess the behaviour and resulting integration is presented. Finally, Section~\ref{sec:Conclusions} concludes the paper highlighting the main contributions that the work described in this paper is bringing into the existing data management and distribution~ecosystem.

\section{Background}\label{sec:SotA}
% Introduction
Before diving into the proposed technical solution, it is crucial to review the key concepts around which this solution has been developed. Firstly, introducing the Open Data paradigm, followed by the well-known Open Data system \gls{CKAN}, and~concluding with the widely used \gls{DCAT-AP} specification. Afterwards, the~\gls{NGSI-LD} standard main features are described, focusing on the information model itself, the~Smart Data Models initiative and the \gls{API} for accessing data from compliant Context Brokers. Moreover, a~review of related works is also provided in this~section.

\subsection{Open Data~Portals}\label{sec:OpenData}
% Open Data
Since the introduction of the first open data portals by the United States government in 2009 and the United Kingdom in 2010, many other countries and organisations have initiated similar open data projects and launched data portals. These initiatives aim to facilitate public access to a diverse array of data available in various formats and spanning a wide range of domains~\cite{yang2013}. This trend aligns with the observations reported in~\cite{umbrich2015}, which noted a continuous growth in the number of datasets and~sources. 

% CKAN
Numerous countries, including a substantial number of EU Member States, have embraced this trend, with~some local governments (e.g., city governments) also participating~\cite{van2014}. Many of these portals utilise the \gls{CKAN}, a~free and open-source data portal platform developed and maintained by Open Knowledge. Consequently, they possess a robust standard \gls{API}, opening up the possibility of amalgamating their catalogues to establish a unified global entry point for discovering and utilising government data. Furthermore, in~line with its open-source approach, \gls{CKAN} allows the development of extensions that modify or add different functionalities to the default Data Portal. This has led to the existence of extensions developed by the \gls{CKAN} community~\cite{ckan_extensions}.

% DCAT-AP
As for the representation and sharing of data descriptions, the~\gls{DCAT-AP} specification is the most widely used. \gls{DCAT-AP} is a \gls{DCAT} profile that provides a standardised way to describe Catalogues containing Datasets and Data Services. It is also designed to enhance interoperability and facilitate the exchange of metadata across different data portals and catalog systems. The~standard includes elements such as dataset titles, descriptions, keywords, and~distribution information to ensure comprehensive and harmonised metadata representation~\cite{dcat-ap_specification}. 

\subsection{NGSI-LD}
% NGSI-LD and Context Broker
The information model specified by the \gls{NGSI-LD} standard is designed to facilitate interoperability between various entities in a digital ecosystem. The~\gls{CB} is the key component in the \gls{NGSI-LD} architecture. It is responsible for managing context information. Furthermore, it enables temporal queries by providing persistent storage of \gls{NGSI-LD} data. Several architectures are possible depending on the organisation of different \glspl{CB}, with~federation being one of the most commonly adopted. The~federated architecture provides the system with a single point of access to the data through the so-called \gls{FCB}. This \gls{FCB} has different \glspl{CB} registered underneath it. As~such, any request made to the \gls{FCB} via this single point of access is forwarded to its federated \glspl{CB} and the results obtained are aggregated to be provided back to the consumer (i.e., the one that made the request). Hence, as~noted above, the~underlying architecture is hidden from the consumer, exposing only the \gls{FCB} \gls{API}~\cite{NGSI-LD-API}. 

% Smart Data Models
NGSI-LD standard only specifies an abstract information model. For~the actual modelling of specific-domain context, the~Smart Data Models initiative provides a common framework for modelling and describing data entities, attributes and relationships to enable the creation of a digital marketplace of interoperable and replicable smart solutions across multiple sectors. This initiative provides more than 900 \gls{NGSI-LD} compliant data models for describing elements in the Smart Agriculture, Smart Cities and Smart Environment domains, among~others, and~can be considered a de facto standard for information representation in these areas~\cite{smartdatamodels}.

\subsection{Related~Works}
Open data portals today play the role of an interface that creates transparency~\cite{van2020}. However, to~provide these opportunities, open data portals should offer users a wide range of mechanisms to enable them to effectively discover, extract and use data~\cite{klein2018}. Nonetheless, according to~\cite{janssen2017}, greater attention needs to be paid to what the transparency promoted by \gls{ICT} is and how it can be achieved. As~it is introduced in~\cite{peled2015} the actual potential of Open Data is not on the release of any information asset as it creates a modern, electronic ``Tower of Babel'' based on incompatible or poorly compatible catalogues through which multiple agents release~information.

Data service is a critical component for Open Data which guarantees the availability of data to users in the form of structured and machine-readable open datasets. Though, aspects such as usability, quality, and~interoperability should be considered in building such open datasets. The~diversity of datasets usually hampers unlocking the full potential value of data. Interoperability addresses the ability of open data platforms and data services to communicate, exchange and consume data, and~to operate effectively~together.

Therefore, various studies have attempted to systematically delineate different interoperability layers. For~example, the~\gls{EIF}~\cite{EIF}, initiated by the European Commission to promote seamless service interoperability and data flows across European public administrations, identifies interoperability layers (Technical, Semantic, Organizational, and~Legal). Drawing inspiration from the \gls{EIF}, the~\gls{EOSC} Interoperability Framework~\cite{eosc} incorporates these layers to enhance interoperability in the research and science domain, aligning with FAIR principles (Findability, Accessibility, Interoperability, and~Reusability) for scientific data management~\cite{fair}. It's worth noting that while \gls{EIF} provides conceptual modelling, \gls{EOSC} concentrates on research data exchange, lacking the flexibility required for the multifaceted nature of versatile Open Data~portals.

In a different approach, a~framework for federated interoperability proposed by~\cite{labreche2020} integrates graph theory and Model-Driven Engineering to enable dynamic data transformation and integration among heterogeneous relational database systems. In~detail, their proposed solution initially explores original relational databases to identify source and target data models, creating their corresponding graph representations. Subsequently, it computes the similarity between elements (nodes and edges) in the two graphs to generate a set of transformation rules for mapping source data to the target data structure. This work serves as a preliminary step in aligning existing datasets with the \gls{NGSI-LD} standard information model, enhancing data exploitability by standardising datasets according to a common model, as~proposed in this~paper.

In~\cite{Won2023} authors identified several problems on the \gls{CKAN}-based Open Data portals ranging from data management limitation, to~no real-time features or absence of interconnection standard, and~proposes a solution for improving interconnectivity and data usability expanding the \gls{CKAN} services. However, it still focuses on the harmonisation of datasets metadata rather than on addressing the challenge of contributing to the usability of the data~itself. 

The \gls{ODDM} proposed in~\cite{Sowe2015} was meant to allow building a platform for interdisciplinary research that makes it easy to extract value from the open data by integrating open data from various sources. However, it is again restricted to the theoretical modelling for developing interoperable Open Data portals, but~without a tangible implementation based on well-established standards and best-practices as we are proposing in this~work.

\section{Data Modelling~Analysis}\label{sec:Connector:Phase0}
The key role of the proposed connector is to accomplish data interpretation and adaptation between both domains (i.e., the context data represented as \gls{NGSI-LD} entities, and~the \gls{CKAN} datasets containing such context data, which are described using \gls{DCAT-AP}). In~order to understand the modules that have been specified for making the two domains to interact with each other, it is crucial to understand the relation between the different data models' features. Tables~\ref{tab:tab1}--\ref{tab:tab3} present the bindings between concepts belonging to the different domains involved in the solution that have been elicited after the analysis of the information models used in those~domains.

\begin{table}[H]
    \caption{DCAT-AP/Catalogue Smart Data Model conversion to CKAN~format.\label{tab:tab1}}
    \newcolumntype{C}{>{\centering\arraybackslash}X}
    \begin{tabularx}{\textwidth}{CC}
        \toprule
        \textbf{Smart Data Model} & \textbf{CKAN} \\
        \texttt{\textbf{Catalogue}}        & \texttt{\textbf{Organization}}  \\
        \midrule
        id                        & id                     \\ 
        \arrayrulecolor{black!70}
        \cmidrule[0.1pt]{1-2}
        \multirow{2.3}{*}{title}    & title                  \\
                                  & name                   \\
        \cmidrule[0.1pt]{1-2}
        description               & description            \\
        \cmidrule[0.1pt]{1-2}
        homepage                  & extras:url             \\
        \cmidrule[0.1pt]{1-2}
        publisher                 & N/A                    \\
        \cmidrule[0.1pt]{1-2}
        \arrayrulecolor{black}
        dataset                   & packages               \\
        \bottomrule
    \end{tabularx}
\end{table}
%\unskip

Identifying the data to be represented was the preliminary step in the analysis. In~this architecture, the~information that flows among the components is not the data points themselves, but~rather the description of datasets. In~other words, descriptive metadata about a collection of data. As~mentioned in Section~\ref{sec:OpenData}, \gls{DCAT-AP} is the reference specification for the exchange of information about catalogues of datasets and data services. Hence, in~our solution we adopted it as a baseline data~model.

\begin{table}[H] 
    \caption{\textls[-25]{DCAT-AP/Dataset
 Smart Data Model conversion to \gls{CKAN} format and \gls{DCAT-AP} specification}. \label{tab:tab2}}

    \newcolumntype{C}{>{\centering\arraybackslash}X}
    \begin{tabularx}{\textwidth}{CCC}
        \toprule
        \textbf{Smart Data Model} & \textbf{CKAN} & \textbf{DCAT-AP} \\
        \texttt{\textbf{Dataset}}               & \texttt{\textbf{Package}}  & \texttt{\textbf{dcat:Dataset}}                                   \\
        \midrule

        id                             & id                     & N/A                                                \\
        \arrayrulecolor{black!70}
        \cmidrule[0.1pt]{1-3}
        description                    & notes                  & dct:description                                    \\
        \cmidrule[0.1pt]{1-3}
        \multirow{2.3}{*}{title}         & title                  & dct:title                                          \\
                                       & name                   & N/A                                                \\
        \cmidrule[0.1pt]{1-3}
        
          accessRights                   & extras:access\_rights  & dct:accessRights                                   \\
        \cmidrule[0.1pt]{1-3}
        creator                        & author                 & N/A                                                \\
        \cmidrule[0.1pt]{1-3}
        distribution                   & resources              & dcat:distribution                                  \\
         \cmidrule[0.1pt]{1-3}
        
         keyword                        & tags                   & dcat:keyword                                       \\
        \cmidrule[0.1pt]{1-3}

%% \bottomrule
% \end{tabularx}
%%\end{adjustwidth}
%\end{table}
%
%\begin{table}[H]\ContinuedFloat
%\small
%\caption{{\em Cont.}}
%
%     \newcolumntype{C}{>{\centering\arraybackslash}X}
%    \begin{tabularx}{\textwidth}{CCC}
%        \toprule
%        \textbf{Smart Data Model} & \textbf{CKAN} & \textbf{DCAT-AP} \\
%        \texttt{\textbf{Dataset}}               & \texttt{\textbf{Package}}  & \texttt{\textbf{dcat:Dataset}}                             \\
%        \midrule      

        landingPage                    & url                    & dcat:landingPage                                   \\
        \cmidrule[0.1pt]{1-3}
        language                       & extras:language        & dct:language                                       \\
        \cmidrule[0.1pt]{1-3}
        license                        & license\_id            & N/A                                                \\
        \cmidrule[0.1pt]{1-3}
        publisher                      & owner\_org             & dct:publisher      \\
        \cmidrule[0.1pt]{1-3}
        spatial                        & extras:spatial         & dct:spatial                                        \\
        \cmidrule[0.1pt]{1-3}
        temporal                       & extras:temporal\_start & dct:temporal  \\
        \cmidrule[0.1pt]{1-3}
        theme                          & extras:theme           & dcat:theme                                         \\
        \cmidrule[0.1pt]{1-3}
        hasVersion                     & extras:has\_version    & N/A                                                \\
        \cmidrule[0.1pt]{1-3}
        versionNotes                   & extras:version\_notes  & owl:versionInfo                                    \\
        \cmidrule[0.1pt]{1-3}
        dataProvider                   & maintainer             & dcat:contactPoint  \\
        \cmidrule[0.1pt]{1-3}
        \multirow{2.2}{*}{dateCreated}   & metadata\_created      & N/A                                                \\
                                       & extras:issued          & dct:issued                                         \\
        \cmidrule[0.1pt]{1-3}
        \arrayrulecolor{black}
        \multirow{2.2}{*}{dateModified}  & metadata\_modified     & N/A                                                \\
                                       & extras:modified        & dct:modified                                       \\
        \bottomrule
    \end{tabularx}
\end{table}
%\unskip

Since the cornerstone of the architecture lies in the \gls{NGSI-LD} environment, the~data models belonging to the \gls{DCAT-AP} subject~\cite{dcat-ap_subject} of the Smart Data Models initiative, which adapt this specification to the \gls{NGSI-LD} context, have been used. In~particular, the~Catalogue, Dataset and Distribution models, which directly adapt these terms from the \mbox{\gls{DCAT-AP}} v2.1.1 specification~\cite{dcat-ap_specification}.

The next data format found in the architecture is the \gls{CKAN} format. It consists of \gls{JSON} objects with specific structures and properties according to the defined term. The~terms Catalogue, Dataset and Distribution, which are defined in \gls{DCAT-AP}, respectively become Organization, Package/Dataset in the \gls{CKAN} specification.

Last but not least, to~ensure maximum interoperability of the whole system as Open Data, the~information (descriptions) provided in \gls{CKAN} shall be made available in \mbox{\gls{DCAT-AP}} format, serialised in \gls{RDF} documents. This requires a final transformation between the \gls{CKAN} format and \gls{DCAT-AP}. 

Note that the transformation presented in Table~\ref{tab:tab1} only goes up to the \gls{CKAN} level due to a module designed in the architecture, as~explained in Section~\ref{sec:Connector:additional}. This module focuses solely on exporting Datasets and Distributions to \gls{DCAT-AP}.

\begin{table}[H] 
    \caption{\textls[-45]{DCAT-AP/Distribution
 Smart Data Model conversion to \gls{CKAN} format and \gls{DCAT-AP} specification}. \label{tab:tab3}}
    \newcolumntype{C}{>{\centering\arraybackslash}X}
    \begin{tabularx}{\textwidth}{CCC}
        \toprule
        \textbf{Smart Data Model} & \textbf{CKAN} & \textbf{DCAT-AP} \\      
        \texttt{\textbf{Distribution}}          & \texttt{\textbf{Resource}}      & \texttt{\textbf{dcat:Distribution}}                         \\
        \midrule
        description                    & description            & dct:description                                    \\
        \arrayrulecolor{black!70}
        \cmidrule[0.1pt]{1-3}
        title                          & name                   & dct:title                                          \\
        \cmidrule[0.1pt]{1-3}
        \multirow{2.3}{*}{accessUrl}     & url                    & N/A                                                \\
                                       & access\_url            & dcat:accessURL                                     \\
        \cmidrule[0.1pt]{1-3}
        availability                   & N/A                    & N/A                                                \\
        \cmidrule[0.1pt]{1-3}
        byteSize                       & size                   & dcat:byteSize                                      \\
        \cmidrule[0.1pt]{1-3}
        downloadURL                    & download\_url          & dcat:downloadURL                                   \\
        \cmidrule[0.1pt]{1-3}
        license                        & license                & dct:license                                        \\
        \cmidrule[0.1pt]{1-3}
        mediaType                      & mimetype               & dcat:mediaType                                     \\
        \cmidrule[0.1pt]{1-3}
        rights                         & rights                 & dct:rights                                         \\
        \cmidrule[0.1pt]{1-3}
        dateCreated                    & created                & dct:issued                                         \\
        \cmidrule[0.1pt]{1-3}
        dateModified                   & last\_modified         & dct:modified                                       \\
        \cmidrule[0.1pt]{1-3}
        \arrayrulecolor{black}
        format                         & format                 & dct:format                                         \\
        \bottomrule
    \end{tabularx}
\end{table}

\section{NGSI-LD to CKAN~Connector}\label{sec:Connector} 
\begin{figure}[H]
%    \centering
    \includegraphics[height=2.3cm]{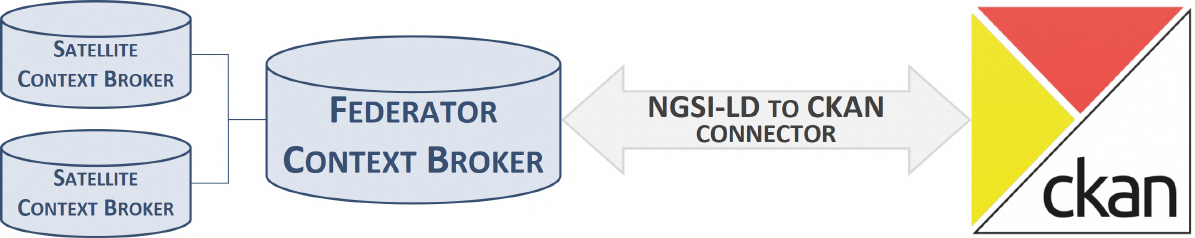}
    \caption{High-level solution setup. \label{fig:fig1}}
\end{figure}

This section presents the proposed solution to address the problem stated in the article: the interworking between the \gls{NGSI-LD} and the Open Data worlds. Figure~\ref{fig:fig1} shows the baseline scenario including the two key components from each of the two worlds that the implemented connector has to make to interwork. The~\gls{NGSI-LD} world is presented on the left side through a set of \glspl{CB} in a federation setup, whereas the Open Data world is depicted on the right side via a \gls{CKAN} instance. The~solution proposed takes place between these two components, along with some other processes needed in order to automate the whole pipeline. These additional modules will be further detailed~below. 

As it was presented in Section~\ref{sec:Connector:Phase0} the communication between a \gls{CB} and an Open Data platform as \gls{CKAN} is not straightforward, since the \gls{CB} stores actual data while the Open Data Portal stores the description of the datasets. Therefore, the~main role of the connector that has been designed and implemented is to create the descriptions of the datasets available through the \gls{CB}, so that \gls{CKAN} is able to consume these descriptions. The~sought-after full connection between these two paradigms, \gls{NGSI-LD} and Open Data, is then ensured by the components and processes of this~architecture.

To this end, we have defined a three-stepped process: firstly, an~understandable description of the data for the \gls{CKAN} to consume has to be generated; secondly, a~harvesting process that imports the descriptions into the \gls{CKAN} instance has to be triggered; and finally, it is necessary to perform the RDF-compliant serialisation for the generated descriptions to be consumed by the \gls{CKAN}-based portals.

\subsection{Phase 1-Creation of~Descriptions}
\begin{figure}[H]
%    \centering
    \includegraphics[height=2.1cm]{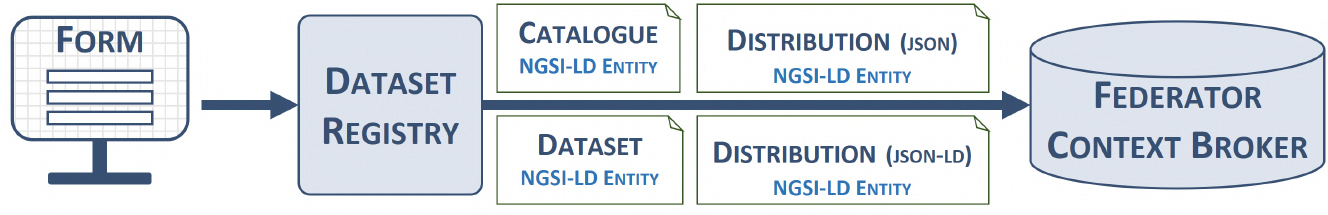}
    \caption{Component architecture of the initial stage. \label{fig:fig2}}
\end{figure}

The initial stage entails the representation of the data. Figure~\ref{fig:fig2} represents the components involved in this first phase, comprising the connection between the user and the \gls{FCB}. That is, the~creation of comprehensive descriptions of the~data.

These descriptions are represented using the \gls{DCAT-AP} subject~\cite{dcat-ap_subject} from the Smart Data Models initiative~\cite{smartdatamodels}, more specifically Catalogue, Dataset and Distribution data models as discussed in Section~\ref{sec:Connector:Phase0}. By~means of these three terms a comprehensive description of a data collection can be made, ranging from the proprietary organisation (Catalogue), through the description of the dataset itself (Dataset), and~finally to the physical representations in different formats of the access to the data in question (Distributions).

The approach followed for the definition of datasets involves grouping data (\gls{NGSI-LD} entities) by entity type (property \texttt{type}). Additionally, two Distributions per Dataset are provided. One using raw \gls{JSON} format, and~the second one employing its linked-data flavour (i.e., \gls{JSON-LD}). Thus, the~proposal is to have a Catalogue that identifies the \gls{FCB} owner, a~Dataset for each data type, and~two Distributions for each Dataset. According to the \gls{DCAT-AP} guidelines~\cite{dcat-ap_specification}, customisation of numerous parameters requires domain knowledge, including descriptions of the elements themselves, location (e.g., Spain, France) and dataset themes (e.g., Environment, Energy, Health). It is strongly recommended that these parameters take one of the available values of the EU Vocabularies in~\cite{publications_europa}, so that they have standard values defined by an external and reliable~institution. 

Taking this into account, a~web-based form (cf. Figure~\ref{fig:fig2}) has been developed as the first element of this initial stage. The~aim of this form is to facilitate to the user the generation of descriptions including these customisable attributes. Fields implicated are listed~below:
\begin{itemize}
    \item \texttt{Type}: https://smartdatamodels.org/dataModel.\textit{<subject>}/\textit{<type>}
    \item \texttt{Description}: \textit{<dataset full description>}
    \item \texttt{Creator}: \textit{<comma-separated list of entities owning or sources of the data>}
    \item \texttt{Provider}: \textit{<comma-separated list of entities publishing data>}
    \item \texttt{Data Type Topic}: \textit{<checkbox list with multiple choice (e.g., Environment, Energy, Health)>}
    \item \texttt{Access Rights}: \textit{<dropdown list (Publich, Restricted, Private)>}
    \item \texttt{Language}: \textit{<dropdown list (English, Spanish, German, French)>}
    \item \texttt{Location}: \textit{<checkbox list with multiple choice (e.g., Austria, Portugal, Spain)>}
    \item \texttt{Keywords}: \textit{<comma-separated list of related concepts>}
\end{itemize}

The next component involved in the architecture is the Dataset Registry module, whose source code is available in~\cite{dataset-registry}. This module generates the \gls{NGSI-LD} entities and injects them into the \gls{FCB}, which is the last element depicted in the initial stage architecture. More precisely, the~Dataset Registry module is responsible for performing the transformation between the output of the form (plain text or \gls{JSON}, depending on the design) and the defined \gls{NGSI-LD} entities. Such transformation exploits the principles of linked data provided by the \gls{NGSI-LD} standard, building relationships between the different entities to generate the tree structure inherent in \gls{DCAT-AP}. Once these entities are generated, in~this case four of them (Catalogue, Dataset, Distribution with \gls{JSON} format and Distribution with \gls{JSON-LD} format), the~component injects them into the \gls{FCB}, making them available from the platform's single access and ready for the next step in the~chain. 

In addition to the fields that directly convey the user's information through the form, there exist certain parameters within the Distributions data models that require some further elaboration. These parameters pertain to the \gls{URL} that enable access to and downloading of the described information. As~mentioned earlier, the~approach taken for clustering data is based on data type (or entity type), therefore the straightforward method of access via the \gls{FCB} through the \gls{NGSI-LD} \gls{API}~\cite{NGSI-LD-API} would be with the resource \texttt{/ngsi-ld/v1/entities/?type=<entity\_type>}. This resource provides all entities of type \texttt{<entity\_type>}, in~line with the Dataset creation strategy. However, rather than the final value of these parameters (\texttt{accessUrl} and \texttt{downloadURL}) being direct links to the context broker, as~described in the section below, an~intermediate module has been developed as a proxy that is responsible for redirecting the~requests. 

\subsection{Phase 2-Publication of Data in CKAN~Instance}
The previous section left the platform in the following state: entities Catalogue, Dataset and Distributions are stored in the \gls{FCB}, all describing the information stored in the federated \glspl{CB}. Therefore, the~next step is the transition to Open Data, for~which a \gls{CKAN} instance is employed. This \gls{CKAN} instance provides a standard representation of the information through a widespread open data portal that is managed by the overall platform~owner. 

\begin{figure}[H]
%    \centering
    \includegraphics[height=3.4cm]{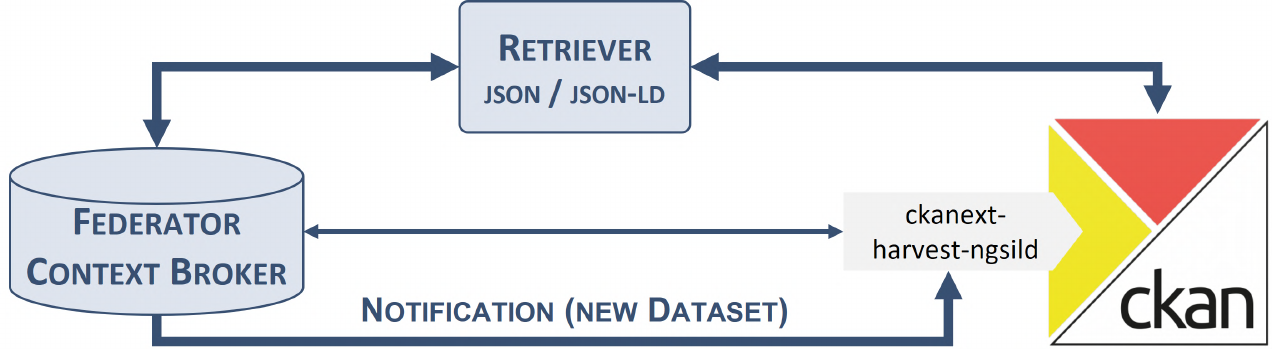}
    \caption{Component architecture of the second stage. \label{fig:fig3}}
\end{figure}

Figure~\ref{fig:fig3} depicts an overview of the components involved in this second phase as well as the interaction between them. As~can be seen, the~platform includes the \gls{FCB}, a~\gls{CKAN} instance with a submodule as an extension, and~the Retriever module, which will be discussed~later.

Focusing on the \gls{CKAN} submodule, it has been already mentioned in Section~\ref{sec:SotA} that \gls{CKAN} allows the development of extensions, so that its functionality and features can be extended for the convenience of the user~\cite{ckan_extensions_documentation}. Taking this into account, this extension submodule called \texttt{ckanext-harvest-ngsild} has been developed (source code available at~\cite{ckanext-harvest-ngsild}), and~is responsible for harvesting the description entities (Catalogue, Dataset and Distribution) stored in the \gls{FCB} and injecting them into the deployed \gls{CKAN} instance. 

The extension module subscribes to Dataset entities in the \gls{FCB}, so that each time a new entity is injected through the Dataset Registry module a notification containing that entity is received. By~means of this information, \texttt{ckanext-harvest-ngsild} transforms both the Catalogue and Dataset, as~well as the associated Distributions from Smart Data Models into the corresponding \gls{CKAN} format (organisation, package and resources) following the mapping presented in Tables~\ref{tab:tab1}--\ref{tab:tab3}, and~injects them into the \gls{CKAN} portal. As~discussed at the end of the previous section, some properties as \texttt{accessUrl} and \texttt{downloadURL} point to the endpoints made accessible by an intermediate module, the~so-called Retriever in Figure~\ref{fig:fig3}. This module is intended to provide the data stored in the federated \glspl{CB} and available through the \gls{FCB} in different formats. Therefore, not only does the Retriever module act as a reverse proxy but it also transforms the data into multiple representations. The~\gls{CKAN} extension provides three additional endpoints to the \gls{CKAN} portal \gls{URL}. These~are:
\begin{itemize}       
    \item \texttt{/ngsi-ld/subscribe}: this endpoint aims to create the subscription to the \gls{FCB} that has just been explained. This subscription, that follows the entity pattern defined in~\cite{NGSI-LD-API}, includes the \texttt{/ngsi-ld/notifications} endpoint as the callback for the notifications. 
    \item \texttt{/ngsi-ld/unsusbcribe}: this endpoint is responsible for unsubscribing from the \gls{FCB}, stopping the reception of notifications triggered by the registration of new entities of type Dataset.
    \item \texttt{/ngsi-ld/notifications}: this endpoints corresponds to the \gls{URL} resource that receives the notifications from the \gls{FCB} to which the module is subscribed. When a notification arrives, it triggers the transformation to \gls{CKAN} format and the creation (or update) of the package (dataset) together with its resources (distributions) in the \gls{CKAN} portal.
\end{itemize}

It is worth noting that these endpoints are restricted to the system administrator, due to the necessity of specific parameters in the requests. These parameters refer to the \gls{CKAN} user performing the request along with an API Token to authenticate against the \gls{CKAN}~platform. 

As discussed previously, two Distribution entities are generated for each Dataset, describing the access to the data in \gls{JSON} and \gls{JSON-LD} format respectively. It has been highlighted that the \texttt{accessUrl} and \texttt{downloadURL} fields are essential to the architecture as they specify the \gls{URL} for accessing and downloading the described data. This \gls{URL} typically points directly to the resources offered by the \gls{FCB}, but~due to limitations in the use of headers required when making requests, it is necessary to generate an intermediate module, called the Retriever. One of these limitations refers to the representation formats. The~\gls{FCB} can provide the stored data in two formats: \gls{JSON} and \gls{JSON-LD}. \textls[-25]{To~enable this option, the~\gls{HTTP} header \texttt{Accept} (i.e., \texttt{Accept:~application/json}) must be} specified. Given this aspect, the~Retriever module is developed (source code available at~\cite{retriever}) and incorporates two~endpoints:
\begin{itemize}       
    \item \texttt{/retriever/realtime/\_\_<url\_type>\_\_.<format>}: this endpoint is intended for real time data requests, i.e.,~the last instance recorded in the \gls{CB}. The~\texttt{<url\_type>} field refers to the complete \gls{URL} path that describes the type of entity being requested and the \texttt{<format>} field refers to the format in which information has to be retrieved. For~example, \texttt{\seqsplit{/retriever/realtime/\_\_https://smartdatamodels.org/dataModel.Parking/ParkingSpot\_\_.jsonld}} will provide the latest values recorded for the entity type ParkingSpot that belongs to the Parking subject~\cite{parkingspot} in \gls{JSON-LD} format.
    \item \textls[-15]{\texttt{\seqsplit{/retriever/temporal/\_\_<url\_type>\_\_.<format>?<temporal\_unit>=<value>}}, where} \texttt{\seqsplit{temporal\_unit = ["year", "months", "weeks", "days", "hours"]}}: this endpoint allows to perform requests with temporal context, i.e.,~making use of the temporal storage provided by the \gls{CB}. The~\texttt{<url\_type>} and \texttt{<format>} fields remain the same as in the previous scenario. Nonetheless, there exist the possibility of adding a query parameter, \texttt{<temporal\_unit>}, which indicates the time unit to be used, and, its value (\texttt{<value>}). The~possibilities currently deployed for the time unit are those discussed above: years, months, weeks, days and hours. For~example, \texttt{\seqsplit{/retriever/temporal/\_\_https://smartdatamodels.org/dataModel.Parking/ParkingSpot\_\_.json?days=5}} will provide the stored values of the last 5 days for the entity type Parking/ParkingSpot~\cite{parkingspot} in \gls{JSON} format.
\end{itemize}

Once the Retriever receives the request, it transfers the necessary parameters to an \gls{NGSI-LD} Query to obtain the values demanded in the first request. This transformation looks like the~following:
\begin{itemize}       
    \item Request sent to Retriever: \\
    \texttt{GET /retriever/realtime/\_\_<url\_type>\_\_.<format>}
    \item Request generated internally in Retriever and sent to \gls{FCB}: \\
    \texttt{GET  /entities?type=<url\_type>}\\
    \textls[-25]{\texttt{Accept:~application/<format>}}
\end{itemize}

Hence, this module is used to supply the \glspl{URL} for accessing the data, with~the appropriate parameters in each individual case. This results in the completion of the Retriever module. Its purpose is twofold: to hide the \gls{FCB} from the outside world (thus protecting against the possibility of malicious requests such as \gls{HTTP} POST or PUT), and~to allow data to be retrieved in multiple formats by means of Open Data~portals. 

\subsection{Phase 3-DCAT~Serialisation} \label{sec:Connector:additional}
This is an additional step towards compliance with the \gls{DCAT-AP} standard beyond Smart Data Models. Essentially, its purpose is to export the data descriptions stored in \gls{CKAN} into \gls{RDF} documents serialised using \gls{DCAT}. Despite \gls{CKAN}'s popularity, this approach ensures maximum interoperability. To~achieve this objective, a~new \gls{CKAN} extension has been developed to transform the descriptions of Organisations, Datasets and Resources 
via the mapping relationships shown in Tables~\ref{tab:tab2} and~\ref{tab:tab3} into \gls{RDF} documents compliant with \gls{DCAT-AP} v2.1.1.

\begin{figure}[H]
%    \centering
    \includegraphics[height=2.3cm]{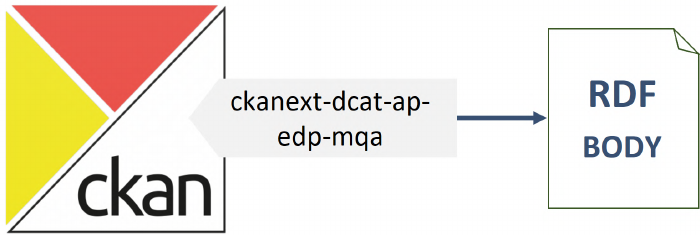}
    \caption{Component architecture of the additional step towards full interoperability. \label{fig:fig4}}
\end{figure}

To this end, the~resulting architecture is shown in Figure~\ref{fig:fig4}, depicting the new extension developed for \gls{CKAN} and its~output. 

\vspace{-3pt}

\textls[-15]{This extension, \texttt{ckanext-dcat-ap-edp-mqa}, uses the \texttt{ckanext-dcat} extension~\cite{ckanext-dcat} as a starting point, which already performs a first transformation between both worlds (\gls{CKAN} and \gls{DCAT-AP}). This base extension enables certain endpoints in the \gls{CKAN} instance, the~most important being \texttt{\seqsplit{https://<ckan-instance-host>/dataset/<dataset-id>.<format>}}.} Via this endpoint, \texttt{<dataset-id>} dataset is exported into the specified \gls{RDF} serialisation format (\texttt{<format>}). The~format parameter is compatible with four values: RDF/XML (\texttt{rdf} or \texttt{xml}, application/rdf+xml), Turtle (\texttt{ttl}, text/turtle), Notation3 (\texttt{n3}, text/n3), and~JSON-LD (\texttt{jsonld}, application/ld+json). Adding the \texttt{profiles} parameter allows customisation on how values defined in \gls{CKAN} are mapped to the \gls{RDF} graph. Moreover, two profiles are provided within this base extension: \texttt{euro\_dcat\_ap\_2} and \texttt{euro\_dcat\_ap}. The~latter is fully compatible with \gls{DCAT-AP} v1.1.1, while the former is partially compatible with \gls{DCAT-AP} v2.1.0. Meanwhile, the~extension proposed in this work, \texttt{ckanext-dcat-ap-edp-mqa}, generates a new profile called \texttt{dcat-ap-edp-mqa} which is fully compatible with \gls{DCAT-AP} v2.1.1, through the transformation presented in Tables~\ref{tab:tab2} and~\ref{tab:tab3}. Thus, using the endpoint shown above by adding the \texttt{profiles} parameter as follows: \texttt{\seqsplit{/dataset/<dataset-id>.<format>?profiles=dcat\_ap\_edp\_mqa}}, the~\gls{RDF} graph of the dataset \texttt{dataset-id} is obtained, fully compatible with the \gls{DCAT-AP} v2.1.1 standard. The~source code of the \texttt{ckanext-dcat-ap-edp-mqa} extension is available at~\cite{ckanext-dcat-ap-edp-mqa}.

\section{Validation Scenario~Implementation}\label{sec:Validation}
This section outlines the specific use case that has been employed for the implementation and validation of the modules described above. The~NGSI-LD-based domain used for the validation was the \gls{DET} implemented and deployed within the \gls{SALTED} project~\cite{Sanchez2023}. This toolchain harmonises and enriches heterogeneous data by applying the principles of linked data and semantics, with~the aim of achieving the so-essential interoperability. Moving on to the \gls{CKAN} domain, the~\gls{SALTED} project has also deployed a dedicated \gls{CKAN} instance~\cite{ckan_salted} that represents the data stored in the \glspl{CB} hosted on the platform. Additionally, the~\gls{EDP} constitutes a final consumer of the data exposed in \gls{CKAN}~\cite{edp}. The~\gls{EDP}~\cite{edp_documentation} is the central point of access to European Open Data. It is the preferred tool for European initiatives to publish their data given its main objectives, among~which its mission is focused on giving open access, high-quality and available data within the European Union. Therefore, it is a reliable source of information and helps with the open dissemination of~data.

\begin{figure}[H]
%    \centering
    \includegraphics[height=4cm]{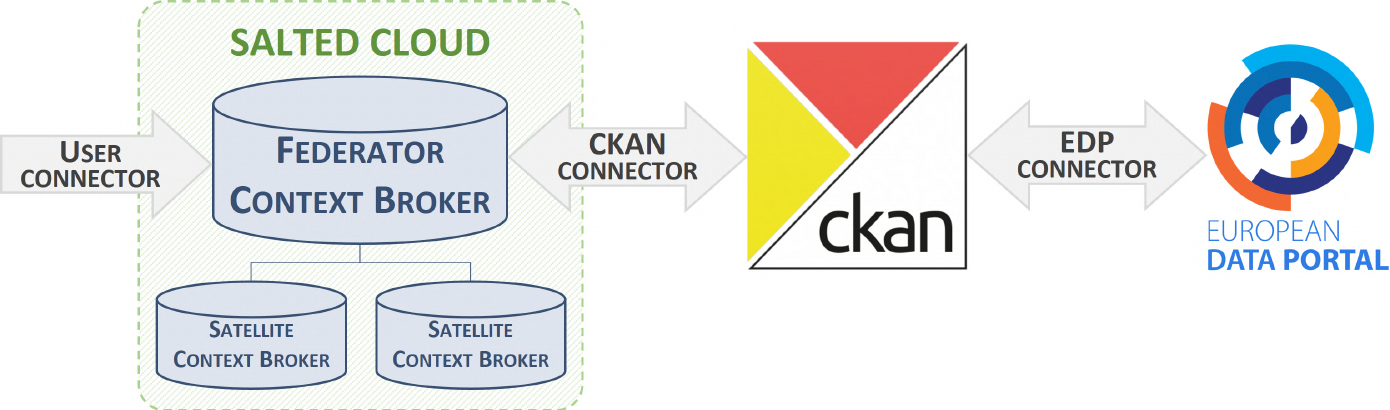}
    \caption{Architecture of the use case with the set of connectors. \label{fig:fig5}}
\end{figure}

Figure~\ref{fig:fig5} illustrates the final architecture of the use case, incorporating the connectors described so far. The~key components are the federated architecture of \glspl{CB}, which exposes the data enriched within the \gls{DET} deployed in \gls{SALTED} (and hosted in the \gls{SALTED} Cloud) through its \gls{NGSI-LD} \gls{API}; the \gls{SALTED} \gls{CKAN}-based Open Data portal, which describes the datasets available; and finally the \gls{EDP} with the \gls{SALTED} data~catalogue.

The integration of \gls{EDP} as a consumer of the information stored in \gls{CKAN} makes it necessary to meet certain requirements concerning the description of the datasets (Catalogue, Datasets and Distribution entities). The~\gls{MQA} procedure~\cite{mqa}, which the Datasets' descriptions undergo once they are published in this Open Data portal, sets out these conditions. As~the central access point to European data, they ensure that the information published through this organisation is of high quality, or~at least that this quality is labelled so that any user seeking to discover and use information is aware of what they are~consuming. 

These enforced \gls{MQA} requirements address these aspects~\cite{mqa}: Findability, using metrics that help people and machines find datasets; Accessibility, using metrics that describe whether access to data is guaranteed; Interoperability, using metrics that determine whether a distribution is considered interoperable; Reusability, using metrics that check whether data is reusable; and Contextuality, using metrics that provide more context to the user. Therefore, the~\gls{MQA} process has been considered in the development of the key~components.

The first component involved is the form, located within the so-called User Connector, which represents the first phase of the workflow. This form has been designed, as~mentioned above, to~allow the user to customise all those parameters of the \gls{DCAT-AP} standard that require some domain knowledge about the data to be described. Likewise, it also sets out to use the default values provided by~\cite{publications_europa}, which are  those checked by the \gls{EDP} in order to guarantee the quality dimensions defined in terms of Data Type Topic, Access Rights, Language and Location (form fields). 

The Dataset Registry, which is also located within the User Connector, is the second component concerned. This module takes the information received from the form and generates the Catalogue, Dataset and Distribution entities. Besides~integrating this information, this module adds additional details to achieve more complete descriptions and thus score higher in the \gls{EDP} \gls{MQA} process. 

Moving on to the next stage, there is the \gls{CKAN} Connector, which brings together the developed extension and the Retriever module. In~terms of the latter, the~addition of the \gls{EDP} at the end of the workflow has been taken into account while generating the \gls{URL} resources for the data access and download properties, as~mentioned above. The~\gls{HTTP} HEAD method has been enabled for the requests received in this module, as~the \gls{EDP} uses this method during the \gls{MQA} process to check the accessibility of these resources. Concerning the extension of the \gls{CKAN} instance, this plug-in performs the conversion from \gls{NGSI-LD} to \gls{CKAN} format, recalling that this transformation makes use of the mapping relationships shown in Tables~\ref{tab:tab1}--\ref{tab:tab3}. The~existence of the \gls{EDP} as a consuming entity does not alter its functionality in this case, since the \gls{NGSI-LD} entities are already completed at \gls{DCAT-AP} level.

\begin{figure}[H]
%    \centering
    \includegraphics[height=2.3cm]{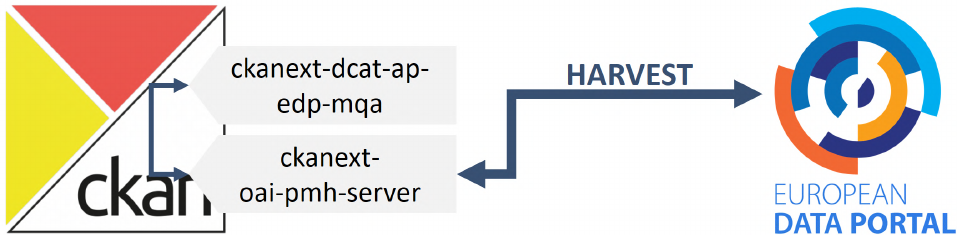}
    \caption{Component architecture of the EDP connector. \label{fig:fig6}}
\end{figure}

The last phase involves the \gls{EDP} connector, which was partially covered in the previous section with the extension \texttt{ckanext-dcat-ap-edp-mqa}. The~behaviour of this extension is verified thanks to the \gls{SHACL} service provided by data.europa.eu~\cite{shacl}, which performs a validation of the \gls{RDF} document body. However, an~additional extension has been developed to ensure that the \gls{CKAN}-stored information is understandable to the \gls{EDP}, by~creating a resource to which the \gls{EDP} can make requests and consume the \gls{DCAT-AP} descriptions provided. Figure~\ref{fig:fig6} presents the architecture of the \gls{EDP} connector, showing the already explained \texttt{ckanext-dcat-ap-edp-mqa} extension along with a new one called \texttt{ckanext-oai-pmh-server}. The~source code of this new extension is available at~\cite{ckanext-oai-pmh-server}. According to the \gls{EDP} documentation~\cite{edp_oaipmh}, its current version of the harvester supports harvesting from an \gls{OAI-PMH}~\cite{oaipmh} compliant source. Consequently, in~this final extension, the~\gls{RDF} descriptions of the Datasets and Distributions are available through the usage of this~protocol. 

Following the \gls{OAI-PMH} documentation~\cite{oaipmh} in combination with the \gls{EDP}~\cite{edp_oaipmh}, the~latter says that it only makes use of the \texttt{ListRecords} verb. This verb is designed to harvest records from a repository. In~other words, it returns a list of records stored in a repository, in~our case all those Catalogues, Datasets and Distributions available~\cite{oaipmh}. The~corresponding \gls{URL} looks like this: \texttt{\seqsplit{https://<ckan-instance-host>/oai?verb=ListRecords\&metadataPrefix=<metadataPrefix>}}. In~successive requests the \texttt{resumptionToken} parameter must be used, in~order to keep track of the elements in the~list.

Thanks to this extension, the~\gls{EDP} can now access the data stored in the \gls{CKAN} and publish it in an open and accessible way. The~\gls{SALTED} Project catalogue in the \gls{EDP} can be seen in~\cite{edp_salted}. 

\begin{figure}[H]
%    \centering
    \includegraphics[height=3.8cm]{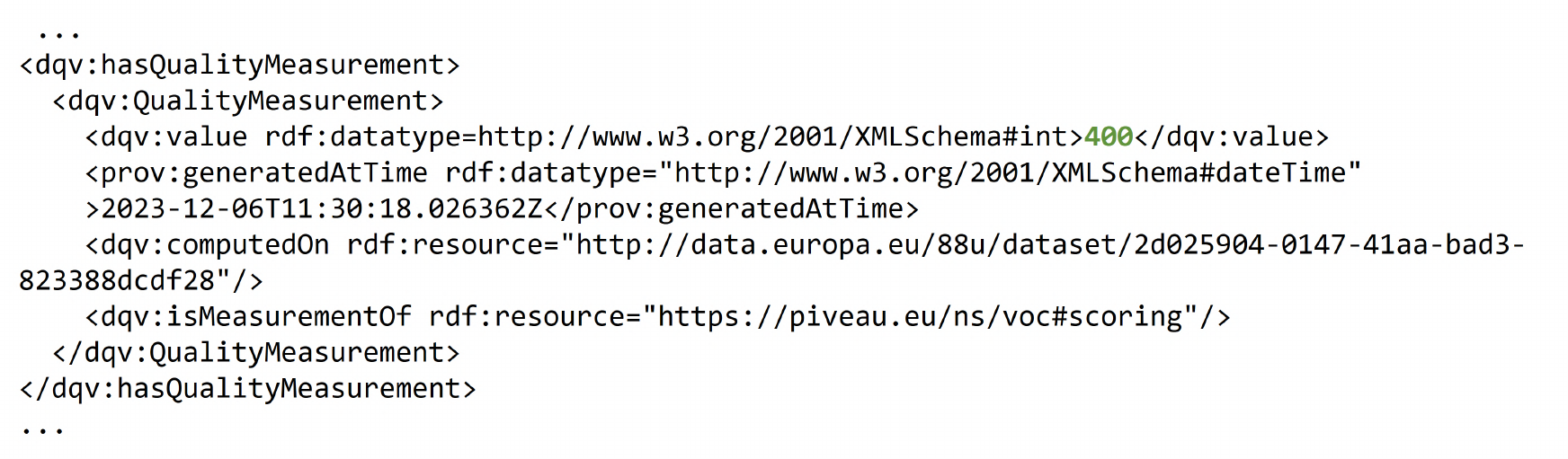}
    \caption{Extract from \gls{DQV} file about the Parking:ParkingSpot dataset~\cite{edp_quality_dqv}. Scoring is highlighted in green. \label{fig:fig8}}
\end{figure}

Quality metrics for a dataset that has undergone the complete workflow are provided in~\cite{edp_quality} and illustrated in Figure~\ref{fig:fig8}. In~\cite{edp_quality} it can be seen that the dataset (Parking:ParkingSpot in particular) scores highly in all metrics except ByteSize, since it was not possible to determine the total size of the dataset due to a design constraint that defined the datasets as real-time values. According to the \gls{EDP} documentation~\cite{mqa}, 405 points is the maximum that can be achieved. In~this particular case, in~the absence of the points related to the ByteSize property, a~total of 400 points is obtained, as~can be seen in Figure~\ref{fig:fig8}. This file extract is sourced from the \gls{DQV} file~\cite{edp_quality_dqv}, encompassing values for each metric and the overall points~computation. 

\vspace{-3pt}

Based on these results, it is clear that the proposed connector architecture achieves the established goals. Starting with the deployment of a \gls{CKAN} instance as an open data portal for the \gls{SALTED} project. This \gls{CKAN} instance is automatically fed through the proposed architecture, so that the open data portal stores data descriptions (Catalogue, Dataset and Distribution) in a way that is compatible with the infrastructure. The~second major result is demonstrated by the interoperability achieved with \gls{DCAT-AP} from the \gls{CKAN} instance itself. For~the third significant result, there is the connection between the \gls{CKAN} instance of the project and the \gls{EDP}, implementing the appropriate interface so that the \gls{EDP} can successfully consume the data, publishing the \gls{SALTED} catalogue on its well-known platform. And~finally, the~fourth milestone is achieved through the \gls{EDP}'s \gls{MQA} process score obtained in each of the published datasets as well as at the catalogue level, placing the project and its open data within the Excellent ranking. 

\section{Conclusions}\label{sec:Conclusions}
% Conclusions
Both \gls{NGSI-LD}-based and \gls{CKAN}-based Open Data are domains constantly expanding and growing nowadays, revealing the potential benefit and opportunity for bridging the gap between them. 
The synergy between the Open Data portals' ability to describe datasets and the \gls{NGSI-LD}'s proficiency in providing the actual data through an interoperable data model not only increases the volume of publicly available information, but~also widens the potential for significant analysis and empowered decision~making.

In this paper we have presented the work carried out to develop a solution consisting of a set of connectors aimed at establishing a real connection between the two domains. The~technological contributions of this work~comprise:
\begin{itemize}
    \item The analysis of both domains, finding out the ways to integrate them.
    \item The analysis of the data models and representation formats encountered throughout the entire pipeline.
    \item The design and development of modules and mechanisms for the completion of the proposed architecture.
    \item The integration of these modules within an established platform in order to validate the toolchain proposed.
    \item The integration of these modules with the widely known \gls{EDP} in order to enrich the depth of the public information available through this portal.
\end{itemize}

All in all, the~proposed architecture consists of nine components: Form, to~enable user interaction when describing datasets; Dataset Registry, to~perform the transformation of form input to \gls{NGSI-LD}; Context Brokers architecture, for~the storage and provision of context information; \gls{CKAN}, as~the project's/initiative's personal open data portal; three extensions to \gls{CKAN}, to~extend the functionalities of that platform in terms of data injection, data export in different formats and provision of an interface for data consumption; Retriever, to~hide the Context Brokers architecture from the outside; and finally, the~\gls{EDP}, as~the final element for data consumption and publication in its widely acknowledged open data~portal.

Possible extensions of this work are, firstly, pertaining to the developed Retriever module, to~enlarge its functionality in order to allow more representation formats such as \gls{RDF/XML}, Turtle or Notation3. This leads to the second future idea, which is to modify the Dataset Registry to generate Distributions not only in JSON and JSON-LD, but~also in the rest of these formats supported by the Retriever. On the other hand, another possible future evolution of the connector is to avoid the need for deploying the \gls{CKAN} instance, and~directly bridge any \gls{NGSI-LD} platform with the \gls{EDP} without such an intermediate domain. Nevertheless, this would require a thorough comparison between the two alternatives, both with and without \gls{CKAN}, to~determine which is the most appropriate. In~this paper, the~\gls{CKAN} instance has been used due to the fact that the vast majority of current Open Data cases are based on this framework, allowing each organisation to have its own Open Data portal, regardless of the fact that they may later be aggregated into a broader one such as the EDP. This is, precisely, why we opted for this alternative, as~it allows for a more generic~approach.

% \section{Patents}

% This section is not mandatory, but may be added if there are patents resulting from the work reported in this manuscript.

% %%%%%%%%%%%%%%%%%%%%%%%%%%%%%%%%%%%%%%%%%%
\vspace{6pt} 

%%%%%%%%%%%%%%%%%%%%%%%%%%%%%%%%%%%%%%%%%%
%% optional
%\supplementary{The following supporting information can be downloaded at:  \linksupplementary{s1}, Figure S1: title; Table S1: title; Video S1: title.}

%%%%%%%%%%%%%%%%%%%%%%%%%%%%%%%%%%%%%%%%%%
\authorcontributions{Conceptualisation, L.M., J.L. and L.S.; methodology, L.M. and J.L.; software, L.M., J.L., V.G. and P.S.; validation, L.M., V.G. and J.R.S.; investigation, L.M. and J.L.; writing---original draft preparation, L.M., J.L. and L.S.; writing---review and editing, L.M., L.S., J.L. and J.R.S.; visualisation, V.G. and P.S.; supervision, J.L. and L.S.; project administration, L.S.; funding acquisition, L.S. and J.L. All authors have read and agreed to the published version of the~manuscript.}

\funding{{This work has been partially supported by the project SALTED (Situation-Aware Linked heterogeneous Enriched Data) from the European Union’s Connecting Europe Facility program under the Action Number 2020-EU-IA-0274, and by means of the project SITED (Semantically-enabled Interoperable Trustworthy Enriched Data-spaces) under Grant Agreement No. PID2021-125725OB-I00 funded by MCIN/AEI/10.13039/501100011033 and the European Union FEDER.}}

%%\institutionalreview{}

%%\informedconsent{Any research article describing a study involving humans should contain this statement. Please add ``Informed consent was obtained from all subjects involved in the study.'' OR ``Patient consent was waived due to REASON (please provide a detailed justification).'' OR ``Not applicable'' for studies not involving humans. You might also choose to exclude this statement if the study did not involve humans. %Written informed consent for publication must be obtained from participating patients who can be identified (including by the patients themselves). Please state ``Written informed consent has been obtained from the patient(s) to publish this paper'' if applicable.}

\dataavailability{No new data were created or analyzed in this study. Data sharing is not applicable to this article.} 
%% \dataavailability{We encourage all authors of articles published in MDPI journals to share their research data. In this section, please provide details regarding where data supporting reported results can be found, including links to publicly archived datasets analyzed or generated during the study. Where no new data were created, or where data is unavailable due to privacy or ethical restrictions, a statement is still required. Suggested Data Availability Statements are available in section ``MDPI Research Data Policies'' at \url{https://www.mdpi.com/ethics}.} 

% Only for journal Nursing Reports
%\publicinvolvement{Please describe how the public (patients, consumers, carers) were involved in the research. Consider reporting against the GRIPP2 (Guidance for Reporting Involvement of Patients and the Public) checklist. If the public were not involved in any aspect of the research add: ``No public involvement in any aspect of this research''.}

% Only for journal Nursing Reports
%\guidelinesstandards{Please add a statement indicating which reporting guideline was used when drafting the report. For example, ``This manuscript was drafted against the XXX (the full name of reporting guidelines and citation) for XXX (type of research) research''. A complete list of reporting guidelines can be accessed via the equator network: \url{https://www.equator-network.org/}.}

% \acknowledgments{In this section you can acknowledge any support given which is not covered by the author contribution or funding sections. This may include administrative and technical support, or donations in kind (e.g., materials used for experiments).}

\conflictsofinterest{The authors declare no conflict of interest. The~funders had no role in the design of the study; in the collection, analyses, or~interpretation of data; in the writing of the manuscript; or in the decision to publish the~results.} 

\vspace{-3pt}
%%%%%%%%%%%%%%%%%%%%%%%%%%%%%%%%%%%%%%%%%%
%% Optional
%\sampleavailability{Samples of the compounds ... are available from the authors.}

%% Only for journal Encyclopedia
%\entrylink{The Link to this entry published on the encyclopedia platform.}
\abbreviations{Abbreviations}{
The following abbreviations are used in this manuscript:\\

\vspace{-3pt}
\noindent 
% \printglossary[type=\acronymtype,nonumberlist] %% reviewing used acronyms purposes
\begin{tabular}{@{}ll}
API & Application Programming Interface\\
CB & Context Broker\\
CKAN & Comprehensive Knowledge Archive Network\\
DCAT & Data Catalog Vocabulary\\
DCAT-AP & Data Catalog Vocabulary-Application Profile\\
DET & Data Enrichment Toolchain\\
DMS & Data Management System\\
DQV & Data Quality Vocabulary\\
EDP & European Data Portal\\
FAIR & Findability, Accessibility, Interoperability and Reusability\\
FCB & Federator Context Broker\\
HTTP & Hypertext Transfer Protocol\\
ICT & Information and Communication Technology\\
IoT & Internet of Things\\
JSON & JavaScript Object Notation\\
JSON-LD & JavaScript Object Notation for Linked Data\\
MQA & Metadata Quality Assessment\\
NGSI-LD & Next Generation Service Interface with Linked Data\\
OAI-PMH & Open Archives Initiative Protocol for Metadata Harvesting\\
RDF & Resource Description Framework\\
RDF/XML & XML Syntax for RDF\\
SALTED & Situation-Aware Linked heTerogeneous Enriched Data\\
SHACL & Shapes Constraint Language\\
URL & Uniform Resource Locator\\
W3C & World Wide Web Consortium\\
\end{tabular}
}

%%%%%%%%%%%%%%%%%%%%%%%%%%%%%%%%%%%%%%%%%%
%% Optional
%\appendixtitles{no} % Leave argument "no" if all appendix headings stay EMPTY (then no dot is printed after "Appendix A"). If the appendix sections contain a heading then change the argument to "yes".
%\appendixstart
%\appendix
%\section[\appendixname~\thesection]{}
%\subsection[\appendixname~\thesubsection]{}
%The appendix is an optional section that can contain details and data supplemental to the main text---for example, explanations of experimental details that would disrupt the flow of the main text but nonetheless remain crucial to understanding and reproducing the research shown; figures of replicates for experiments of which representative data are shown in the main text can be added here if brief, or as Supplementary Data. Mathematical proofs of results not central to the paper can be added as an appendix.

%\begin{table}[H] 
%\caption{This is a table caption.\label{tab5}}
%\newcolumntype{C}{>{\centering\arraybackslash}X}
%\begin{tabularx}{\textwidth}{CCC}
%\toprule
%\textbf{Title 1}	& \textbf{Title 2}	& \textbf{Title 3}\\
%\midrule
%Entry 1		& Data			& Data\\
%Entry 2		& Data			& Data\\
%\bottomrule
%\end{tabularx}
%\end{table}

%\section[\appendixname~\thesection]{}
%All appendix sections must be cited in the main text. In the appendices, Figures, Tables, etc. should be labeled, starting with ``A''---e.g., Figure A1, Figure A2, etc.

%%%%%%%%%%%%%%%%%%%%%%%%%%%%%%%%%%%%%%%%%%
\begin{adjustwidth}{-\extralength}{0cm}

\reftitle{References}

\PublishersNote{}
\end{adjustwidth}

\begin{thebibliography}{999}

\bibitem[Louren{\c{c}}o et~al.(2017)Louren{\c{c}}o, Piotrowski, and
Ingrams]{lourencco2017}
Louren{\c{c}}o, R.P.; Piotrowski, S.; Ingrams, A.
\newblock Open data driven public accountability.
\newblock {\em Transform. Gov. People Process Policy} {\bf 2017},
{\em 11},~42--57. [\href{http://doi.org/10.1108/TG-12-2015-0050}{CrossRef}]

\bibitem[Noveck(2017)]{noveck2017}
Noveck, B.S.
\newblock Rights-based and tech-driven: Open data, freedom of information, and
the future of government transparency.
\newblock {\em Yale Hum. Rts. Dev. LJ} {\bf 2017}, {\em 19},~1.

\bibitem[Pereira et~al.(2017)Pereira, Macadar, Luciano, and Testa]{pereira2017}
Pereira, G.V.; Macadar, M.A.; Luciano, E.M.; Testa, M.G.
\newblock Delivering public value through open government data initiatives in a
Smart City context.
\newblock {\em Inf. Syst. Front.} {\bf 2017}, {\em 19},~213--229. [\href{http://dx.doi.org/10.1007/s10796-016-9673-7}{CrossRef}]

\bibitem[{Context Information Management (CIM) ETSI Industry Specification
Group (ISG)}(2023{\natexlab{a}})]{NGSI-LD-API}
{Context Information Management (CIM) ETSI Industry Specification Group (ISG)}.
\newblock {NGSI-LD API}.
\newblock Available online:
\url{https://www.etsi.org/deliver/etsi_gs/CIM/001_099/009/01.07.01_60/gs_CIM009v010701p.pdf}
(accessed on {3 January 2024}
).

\bibitem[{Context Information Management (CIM) ETSI Industry Specification
Group (ISG)}(2023{\natexlab{b}})]{NGSI-LD-InformationModel}
{Context Information Management (CIM) ETSI Industry Specification Group (ISG)}. {NGSI-LD Information Model}.  Available online:
\url{https://www.etsi.org/deliver/etsi_gs/CIM/001_099/006/01.02.01_60/gs_CIM006v010201p.pdf}
(accessed on 3 January 2024).

\bibitem[Teixeira et~al.(2023)Teixeira, Sargento, Rito, Lu{\'\i}s, and
Castro]{teixeira2023}
Teixeira, P.; Sargento, S.; Rito, P.; Lu{\'\i}s, M.; Castro, F.
\newblock A Sensing, Communication and Computing Approach for Vulnerable Road
Users Safety.
\newblock {\em IEEE Access} {\bf 2023}, {\em 11},~4914--4930. [\href{http://dx.doi.org/10.1109/ACCESS.2023.3235863}{CrossRef}]

\bibitem[Sotres et~al.(2019)Sotres, Lanza, S{\'a}nchez, Santana, L{\'o}pez, and
Mu{\~n}oz]{sotres2019}
Sotres, P.; Lanza, J.; S{\'a}nchez, L.; Santana, J.R.; L{\'o}pez, C.;
Mu{\~n}oz, L.
\newblock Breaking vendors and city locks through a semantic-enabled global
interoperable internet-of-things system: A smart parking case.
\newblock {\em Sensors} {\bf 2019}, {\em 19},~229. [\href{http://dx.doi.org/10.3390/s19020229}{CrossRef}]

\bibitem[dca()]{dcat-ap_specification}
DCAT Application Porfile for Data Portals in Europe-Version 2.1.1.
\newblock Available online:
\url{https://github.com/SEMICeu/DCAT-AP/blob/b9b20d1d25e6d827754e93af918344a46dc41a1b/releases/2.1.1/dcat-ap_2.1.1.pdf}
(accessed on 18 January 2024).

\bibitem[Yang and Kankanhalli(2013)]{yang2013}
Yang, Z.; Kankanhalli, A.
\newblock Innovation in government services: The case of open data.
\newblock In Proceedings of the Grand Successes and Failures in IT. Public and
Private Sectors: IFIP WG 8.6 International Working Conference on Transfer and
Diffusion of IT, TDIT 2013, Bangalore, India, 27--29 June 2013; Proceedings;
Springer:  {Berlin/Heidelberg, Germany,} 
2013; pp. 644--651.

\bibitem[Umbrich et~al.(2015)Umbrich, Neumaier, and Polleres]{umbrich2015}
Umbrich, J.; Neumaier, S.; Polleres, A.
\newblock Quality assessment and evolution of open data portals.
\newblock In Proceedings of the 2015 3rd International Conference on Future
Internet of Things and Cloud, {Rome, Italy, 24--26 August 2015}; IEEE: {Piscataway, NJ, USA}, 2015; pp. 404--411.

\bibitem[van~der Waal et~al.(2014)van~der Waal, W{\k{e}}cel, Ermilov, Janev,
Milo{\v{s}}evi{\'c}, and Wainwright]{van2014}
van~der Waal, S.; W{\k{e}}cel, K.; Ermilov, I.; Janev, V.; Milo{\v{s}}evi{\'c},
U.; Wainwright, M.
\newblock Lifting open data portals to the data web.
\newblock In {\em Linked Open Data--Creating Knowledge Out of Interlinked Data:
Results of the LOD2 Project}; {Springer: Berlin/Heidelberg, Germany,} 2014; pp. 175--195.

\bibitem[cka()]{ckan_extensions}
{Find CKAN Extensions-CKAN Extensions}.
\newblock Available online: \url{https://extensions.ckan.org/} (accessed on 22 January 2024).

\bibitem[sma()]{smartdatamodels}
Smart {Data} {Models} Initiative--A global Program led by FIWARE Foundation,
TMForum, IUDX and OASC.
\newblock Available online: \url{https://smartdatamodels.org/} (accessed on 17 November 2023).

\bibitem[van Schalkwyk and Ca{\~n}ares(2020)]{van2020}
van Schalkwyk, F.; Ca{\~n}ares, M.
\newblock 10 Open Government Data for Inclusive Development.
\newblock In {\em Making Open Development Inclusive}; {MIT Press: Cambridge, MA, USA,} 2020; p. 251.

\bibitem[Klein et~al.(2018)Klein, Klein, and Luciano]{klein2018}
Klein, R.H.; Klein, D.C.B.; Luciano, E.M.
\newblock Identification of mechanisms for the increase of transparency in open
data portals: An~analysis in the Brazilian context.
\newblock {\em Cad. EBAPE BR} {\bf 2018}, {\em 16},~692--715. [\href{http://dx.doi.org/10.1590/1679-395173241}{CrossRef}]

\bibitem[Janssen et~al.(2017)Janssen, Matheus, Longo, and
Weerakkody]{janssen2017}
Janssen, M.; Matheus, R.; Longo, J.; Weerakkody, V.
\newblock Transparency-by-design as a foundation for open government.
\newblock {\em Transform. Gov. People Process Policy} {\bf 2017},
{\em 11},~2--8. [\href{http://dx.doi.org/10.1108/TG-02-2017-0015}{CrossRef}]

\bibitem[Peled and Nahon(2015)]{peled2015}
Peled, A.; Nahon, K.
\newblock {Towards open data for public accountability: Examining the US and the
UK models. }
\newblock In Proceedings of the iConference 2015, Newport, CA, USA, 24--27 March 2015. 


\bibitem[{European Commission}(2017)]{EIF}
{European Commission}.
\newblock New European Interoperability Framework.
\newblock Available online:
\url{https://ec.europa.eu/isa2/sites/default/files/eif_brochure_final.pdf}
(accessed on 3 January 2024).

\bibitem[Corcho et~al.(2021)Corcho, Eriksson, Kurowski, Ojster{\v{s}}ek,
Choirat, Van~de Sanden, and Coppens]{eosc}
Corcho, O.; Eriksson, M.; Kurowski, K.; Ojster{\v{s}}ek, M.; Choirat, C.;
Van~de Sanden, M.; Coppens, F.
\newblock {\em EOSC Interoperability Framework: Report from the EOSC Executive
Board Working Groups FAIR and Architecture}; Univerza v Mariboru, Fakulteta
za Elektrotehniko, Ra{\v{c}}unalni{\v{s}}tvo in Informatiko: {Maribor, Slovenia,}  2021.

\bibitem[Wilkinson et~al.(2016)Wilkinson, Dumontier, Aalbersberg, Appleton,
Axton, Baak, Blomberg, Boiten, da~Silva~Santos, Bourne, et~al.]{fair}
Wilkinson, M.D.; Dumontier, M.; Aalbersberg, I.J.; Appleton, G.; Axton, M.;
Baak, A.; Blomberg, N.; Boiten, J.W.; da~Silva~Santos, L.B.; Bourne, P.E.;
et~al.
\newblock The FAIR Guiding Principles for scientific data management and
stewardship.
\newblock {\em Sci. Data} {\bf 2016}, {\em 3},~{160018.}  [\href{http://dx.doi.org/10.1038/sdata.2016.18}{CrossRef}] [\href{http://www.ncbi.nlm.nih.gov/pubmed/26978244}{PubMed}]


\bibitem[Labreche et~al.(2020)Labreche, Montarnal, Truptil, Lorca, Weill, and
Adi]{labreche2020}
Labreche, M.; Montarnal, A.; Truptil, S.; Lorca, X.; Weill, S.; Adi, J.P.
\newblock Towards a Framework for Federated Interoperability to Implement an
Automated Model Transformation.
\newblock In Proceedings of the Boosting Collaborative Networks 4.0: 21st IFIP
WG 5.5 Working Conference on Virtual Enterprises, PRO-VE 2020, Valencia,
Spain, 23--25 November 2020; Proceedings 21; Springer:  {Berlin/Heidelberg, Germany,} 2020; pp. 143--152.

\bibitem[Won et~al.(2023)Won, Han, Gil, and Moon]{Won2023}
Won, H.; Han, J.; Gil, M.S.; Moon, Y.S.
\newblock SODAS: Smart Open Data as a Service for Improving Interconnectivity
and Data Usability.
\newblock {\em Electronics} {\bf 2023}, {\em 12}, {1237.}
\newblock {[\href{http://dx.doi.org/10.3390/electronics12051237}{CrossRef}]}

\bibitem[Sowe and Zettsu(2015)]{Sowe2015}
Sowe, S.K.; Zettsu, K.
\newblock Towards an Open Data Development Model for Linking Heterogeneous Data
Sources.
\newblock In Proceedings of the 2015 Seventh International Conference on
Knowledge and Systems Engineering (KSE),  {Ho Chi Minh City, Vietnam, 8--10 October} 2015; pp. 344--347.
\newblock {[\href{http://dx.doi.org/10.1109/KSE.2015.56}{CrossRef}]}

\bibitem[dca()]{dcat-ap_subject}
{DCAT}-{AP} Subject, Smart Data Model, GitHub Repository.
\newblock Available online:
\url{https://github.com/smart-data-models/dataModel.DCAT-AP/tree/master}
(accessed on 17 November 2023).

\bibitem[pub()]{publications_europa}
EU Vocabularies-Publications Office of the European Union.
\newblock Available online:
\url{https://op.europa.eu/en/web/eu-vocabularies/data-catalogue} (accessed on 20 November 2023).

\bibitem[dat()]{dataset-registry}
Dataset Registry Source Code, GitHub Repository.
\newblock Available online:
\url{https://github.com/tlmat-unican/salted-dataset-registry} (accessed on 18 January 2024).

\bibitem[cka({\natexlab{a}})]{ckan_extensions_documentation}
Writing Extensions Tutorial---{CKAN} 2.10.1 Documentation.
\newblock Available online:
\url{https://docs.ckan.org/en/2.10/extensions/tutorial.html#} (accessed on 23 November 2023).

\bibitem[cka({\natexlab{b}})]{ckanext-harvest-ngsild}
Ckanext-Harvest-Ngsild, GitHub Repository.
\newblock Available online:
\url{https://github.com/tlmat-unican/ckanext-harvest-ngsild} (accessed on
18 January 2024).

\bibitem[ret()]{retriever}
Retriever Source Code, GitHub Repository.
\newblock Available online:
\url{https://github.com/tlmat-unican/salted-retriever} (accessed on
\mbox{18 January 2024}).

\bibitem[par()]{parkingspot}
Parking/ParkingSpot Smart Data Model, GitHub Repository.
\newblock Available online:
\url{https://github.com/smart-data-models/dataModel.Parking/blob/master/ParkingSpot/doc/spec.md}
(accessed on 18 January 2024).

\bibitem[cka({\natexlab{a}})]{ckanext-dcat}
Ckanext-Dcat, GitHub Repository.
\newblock Available online: \url{https://github.com/ckan/ckanext-dcat}
(accessed on 17 November 2023).

\bibitem[cka({\natexlab{b}})]{ckanext-dcat-ap-edp-mqa}
Ckanext-Dcat-ap-edp-mqa, GitHub Repository.
\newblock Available online:
\url{https://github.com/tlmat-unican/ckanext-dcat-ap-edp-mqa} (accessed on
18 January 2024).

\bibitem[Sanchez et~al.(2023)Sanchez, Lanza, Santana, Sotres, Gonzalez, Martin,
Solmaz, Kovacs, Dietzel, Summa, Jafari, Minerva, and Crespi]{Sanchez2023}
Sanchez, L.; Lanza, J.; Santana, J.R.; Sotres, P.; Gonzalez, V.; Martin, L.;
Solmaz, G.; Kovacs, E.; Dietzel, M.; Summa, A.;  et~al.
\newblock Data Enrichment Toolchain: A Data Linking and Enrichment Platform for
Heterogeneous Data.
\newblock {\em IEEE Access} {\bf 2023}, {\em 11},~103079--103091.
\newblock {[\href{http://dx.doi.org/10.1109/ACCESS.2023.3317705}{CrossRef}]}

\bibitem[cka()]{ckan_salted}
SALTED Project CKAN.
\newblock Available online: \url{https://ckan.salted-project.eu/} (accessed on 27 November 2023).

\bibitem[edp({\natexlab{a}})]{edp}
\textls[-15]{data.europa.eu-The Official Portal for European Data. Available online: \url{https://data.europa.eu/en} (accessed on
27 November 2023).}

\bibitem[edp({\natexlab{b}})]{edp_documentation}
Documentation of data.europa.eu ({DEU}).
\newblock Available online:
\url{https://dataeuropa.gitlab.io/data-provider-manual/} (accessed on
\mbox{14 November 2023}).

\bibitem[mqa()]{mqa}
Metadata Quality Dashboard-{Documentation} of data.europa.eu ({DEU}).
\newblock Available online:
\url{https://dataeuropa.gitlab.io/data-provider-manual/metadata-quality/#metadata-quality-dashboard/}
(accessed on 17 November 2023).

\bibitem[sha()]{shacl}
SHACL Service, data.europa.eu.
\newblock Available online: \url{https://data.europa.eu/api/mqa/shacl/}
(accessed on 23 November 2023).

\bibitem[cka()]{ckanext-oai-pmh-server}
ckanext-oai-pmh-server, GitHub Repository.
\newblock Available online:
\url{https://github.com/tlmat-unican/ckanext-oai-pmh-server} (accessed on 18 January 2024).

\bibitem[edp()]{edp_oaipmh}
How to rEquest Harvesting-{Documentation} of data.europa.eu ({DEU}).
\newblock Available online:
\url{https://dataeuropa.gitlab.io/data-provider-manual/how-to-publish/request-harvesting/#supported-formats-and-protocols}
(accessed on 28 November 2023).

\bibitem[Lagoze et~al.(2002)Lagoze, Van~de Sompel, Nelson, and Warner]{oaipmh}
Lagoze, C.; Van~de Sompel, H.; Nelson, M.; Warner, S.
\newblock Open {Archives} {Initiative}-{Protocol} for {Metadata} {Harvesting}
-v.2.0.
\newblock Available online:
\url{https://www.openarchives.org/OAI/openarchivesprotocol.html} (accessed on
28 November 2023).

\bibitem[edp({\natexlab{a}})]{edp_salted}
SALTED Project Catalogue-data.europa.eu.
\newblock Available online:
\url{https://data.europa.eu/data/catalogues/salted?locale=en} (accessed on
18 January 2024).

\bibitem[edp({\natexlab{b}})]{edp_quality}
Parking:ParkingSpot Dataset-Quality Metrics-data.europa.eu.
\newblock Available online:
\url{https://data.europa.eu/data/datasets/2d025904-0147-41aa-bad3-823388dcdf28/quality?locale=en}
(accessed on 11 January 2024).

\bibitem[edp({\natexlab{c}})]{edp_quality_dqv}
DQV Metrics File for Parking:ParkingSpot Dataset-data.europa.eu.
\newblock Available online:
\url{https://data.europa.eu/api/hub/repo/datasets/2d025904-0147-41aa-bad3-823388dcdf28.rdf/metrics}
(accessed on 11 January {2024).} 


\end{thebibliography}
\end{document}